\documentclass[prl,twocolumn,showpacs]{revtex4}
\usepackage{graphicx,bm,amssymb,amsmath}

\newcommand{\pdag}{{\phantom{\dagger}}}

\begin{document}

\title{Kondo ``underscreening'' cloud: spin-spin correlations around
a partially screened magnetic impurity}

\author{L\'aszl\'o Borda$^{1}$, Markus Garst$^{2}$, and Johann Kroha$^{1}$}
\affiliation{$^{1}$Physikalisches Institut and Bethe Center for Theoretical Physics, Universit\"at Bonn, Nussallee 12, 
D-53115 Bonn, Germany\\
$^{2}$Institut f\"ur Theoretische Physik, Universit\"at zu K\"oln, 
Z\"ulpicher Str. 77, D-50937 K\"oln, Germany
}

\date{\today}

\begin{abstract}
{
We consider the spatial spin correlations around a partially screened spin-1 magnetic moment in
a metal exhibiting the underscreened Kondo effect. We find that the
underscreening of the impurity spin results in spatial spin correlations that
are more pronounced as compared to the fully screened Kondo effect; their
power-law decay is weaker because of characteristic logarithmic corrections at
large distances. 
The spin correlator also changes sign as a function of distance to the impurity allowing for ferromagnetic correlations between conduction electron spin density and the local moment. The numerical findings are shown to be in agreement with the predictions deriving from an effective ferromagnetic Kondo Hamiltonian.
}
 
\end{abstract}

\maketitle

{\em Introduction} ---
Spatial spin correlations around Kondo impurities -- the Kondo spin screening cloud -- and its possible observability in experimental setups have been studied intensively over the last decades \cite{Hewson,Ishii78,BarzykinAffleck98,Affleck01,Sorenson05,Borda07,AffleckSaleur08}. Most of the work focused on the properties of spin correlations within the spin-$\frac{1}{2}$ Kondo model, which is thought to capture the 
essential low-energy physics of magnetic ions in metals. It exhibits 
complete spin screening and Fermi liquid 
behavior below the Kondo 
temperature $T_K$, two characteristic features that also determine the 
long-distance behavior of the spin correlations. However,
most transition metal and rare earth atoms have a spin magnetic moment 
$S$ greater than $\frac{1}{2}$ in the ground state. Hence, when immersed in a metal
they can potentially form underscreened Kondo impurity systems, even if the 
local moment degeneracy is partially lifted due to crystal field or spin-orbit
splitting. 

The ground state of the spin-$1$ single-channel Kondo model -- the prototype of underscreened 
Kondo models -- is known to be effectively Nozi\`eres Fermi liquid with a decoupled impurity spin-$\frac{1}{2}$. Indeed, a careful analysis of the conduction electron scattering matrix \cite{MehtaZarand05,KollerMeyer05} revealed that the spectrum exhibits a one-to-one correspondence to that of a free Fermi gas. However, the residual ferromagnetic interaction of the Nozi\'eres quasiparticles with the residual impurity spin-$\frac{1}{2}$ results in non-analytic Fermi liquid corrections. 
Although there are well-defined quasiparticles at the Fermi surface, their 
decay rate vanishes with energy in a singular fashion. Correspondingly, 
the low-energy properties of the underscreened Kondo models were termed 
as {\it singular Fermi liquid} behavior \cite{MehtaZarand05}, in particular, 
to contrast it with the genuine non-Fermi liquid characteristics of the 
overscreened Kondo models. The anomalous corrections should affect all 
physical quantities that involve the conduction electron scattering 
matrix like, e.g., conductivity or, as we show below, the long-distance tail of the spin correlations.

Generally, underscreened Kondo physics emerges if the effective number of screening channels that couple to the local moment carrying a spin $S$ is smaller than $2S$. Although the ground state of a Kondo spin in experimental systems is a fully screened singlet in most of the cases, there might be an extended temperature regime with an underscreened moment as the coupling of some of the channels can be very weak. For example, the screening of a magnetic impurity atom with spin $S>\frac{1}{2}$ adsorbed on metallic surfaces is restricted due to the lower symmetry near the surface potentially giving rise to underscreening \cite{orsi,rok}. A realization in quantum dot setups was also proposed \cite{PustilnikGlazman04,PosazhennikovaColeman05}.  If a quantum dot with a spin $S=1$ ground state is asymmetrically attached to leads, the spin is quenched in a two-stage process. The stronger coupled channel first screens half of the spin-$1$ at a temperature $T_{K1}$, and the weaker coupled channel subsequently screens the residual spin-$\frac{1}{2}$ at a lower temperature $T_{K2}$.  For intermediate temperatures, $T_{K2}<T<T_{K1}$, the quantum dot system is governed by the characteristics of the underscreened spin-$1$ Kondo model, and its singular dynamics might be experimentally accessible in this temperature range.

In this work, we focus on the spatial, equal-time spin-correlations of the
conduction electrons around a spin-$1$ impurity, that provide a snapshot of
the screening cloud. In general, the characteristics of the screening cloud in
Kondo models are less well studied than thermodynamic or transport
properties. In the fully screened spin-$\frac{1}{2}$ model, a number of
theoretical results\cite{Ishii78,BarzykinAffleck98,Borda07} 
support the expectation that the spatial spin-correlations exhibit a crossover
at a length scale $\xi_K \equiv \hbar v_F / (k_B T_K)$, where $v_F$ is the
Fermi velocity and $T_K$ is the Kondo temperature\cite{BarzykinAffleck98}; for
a review see Section 9.6 of Ref.~[\onlinecite{Hewson}]. Generally at zero
temperature, spin-density correlations decay as a power-law   with increasing
distance $x$ from the impurity.  For the fully screened spin-$\frac{1}{2}$
impurity, the crossover is accompanied by a change in the power-law exponent;
while at short distances, $x < \xi_K$, spin correlations decay as $1/x^d$,
with $d$ the dimensionality of the conduction electron system, there is a
stronger $1/x^{d+1}$ decay at large distances, $x > \xi_K$ \cite{Ishii78,BarzykinAffleck98,Borda07}. Recently, it was shown that even the charge-density oscillations around the fully screened impurity bear signatures of this crossover \cite{AffleckSaleur08}, albeit with much weaker characteristics. On the experimental side, the detection of this crossover in spatial correlations has proved to be elusive so far. The most promising system for its observation is probably a magnetic impurity placed on a metallic surface subject to an STM study \cite{AffleckSaleur08}.

Comparatively, only little is known about the spatial correlations around a
partially screened magnetic impurity. The question arises as to how the singular Fermi liquid properties of the underscreened spin-$1$ Kondo model modify the crossover at $\xi_K$. The scattering of the electrons off the unscreened residual moment will influence the long-distance behavior of spatial correlations, and
the conduction electron cloud around a partially screened impurity --- the "underscreening" cloud --- will thus differ from its fully screened counterpart. We study this question numerically with the help of an extention of Wilson's numerical renormalization group (NRG) technique \cite{Borda07}. We find that within the numerical accuracy the envelope of the equal-time impurity-spin -- electron-spin correlator is a universal function of $x/\xi_K$. In contrast to the spin-$\frac{1}{2}$ model, its power-law decay obtains now only logarithmic corrections for $x > \xi_K$, and its long-distance tail is thus more pronounced than in the fully screened case. We demonstrate that this numerically observed tail of the  "underscreening cloud" is in agreement with the predictions of an effective ferromagnetic spin-$\frac{1}{2}$ Kondo model describing the scattering off the residual magnetic moment.
 
\begin{figure*}
\includegraphics[width=1.8\columnwidth,clip]{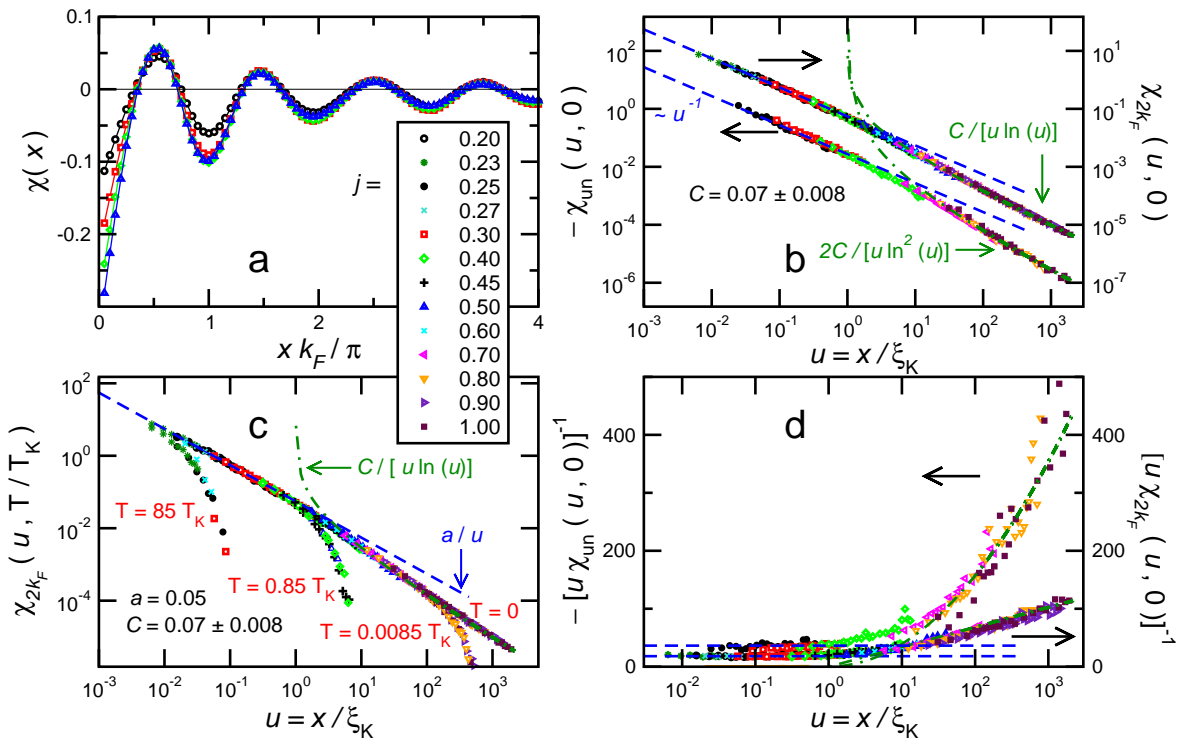}
\caption{(Color online)
Panel a: The equal time spin-spin correlation function, 
$\chi(x,T)=\langle\vec{S}_{\rm imp}\,\vec{s}(x)\rangle$ as a
function of the distance $x$ measured from the $S=1$ impurity for
different values of the Kondo coupling $j=\nu J$. 
Unlike the situation for the completely screened 
Kondo impurity, $\chi$ 
changes sign as a function of distance, and 
the electron spin density close to $x=(n+\frac{1}{2})\frac{\pi}{k_F}$
is aligned, rather than being anticorrelated, with the impurity spin. 
In panel b, we show the amplitude of the $2 k_F$ oscillating and uniform part of $\chi$ at $T=0$ that are universal functions of $x/\xi_K$, see Eq.~(\ref{UniversalAsympt}).
While for short distances both functions decay as $\sim x^{-1}$, they both obtain logarithmic corrections for $x > \xi_K$.
As shown in panel c, the envelope of the oscillating part 
crosses over from $\sim x^{-1}$ to
$\sim 1/[x\ln(x/\xi_K)]$ at 
around the Kondo coherence length, $\xi_K$.
Any finite temperature introduces another length scale,
$\xi_T=\hbar v_F/(k_B T)$ at which the envelope function crosses over 
to an exponential decay.
The uniform part decays faster beyond the Kondo coherence length: it tends to
zero as $\sim 1/[x\ln^{2}(x/\xi_K)]$. For better visualization, we plot
$1/[x\chi(x)]$ as a function of $x/\xi_K$ 
in panel d.
}
\label{fig:equalt}
\end{figure*}

{\em Model and definition of correlation functions} ---
The Hamiltonian of the spin-1 single-channel Kondo model reads
\begin{align} \label{Model}
\mathcal{H} = J \vec{S}_{\rm imp}\vec{s}(0)
+ \sum_{k \sigma} \varepsilon_k c^\dagger_{k\sigma} c^\pdag_{k\sigma}, 
\end{align}
where $\vec{S}_{\rm imp}$ is the impurity spin-1 located at $x=0$ and 
$c^\dagger_{k\sigma}$ ($c^{\pdag}_{k\sigma}$) creates (annihilates)
a conduction electron with momentum $k$ and spin $\sigma$. The impurity spin
is antiferromagnetically ($J>0$) and locally coupled to the electron spin density,
$\vec{s}(x)=\psi_\alpha^\dagger(x)\frac{\vec{\sigma}_{\alpha\beta}}{2}\psi_\beta(x)$, with $\psi(x) = \sum_k e^{i k x} c_{k\sigma}$. We limit ourselves here to the case of one-dimensional conduction electrons, $d=1$. 
A perturbative treatment of the coupling $J$ encounters logarithmic low-energy divergencies that result in a breakdown of perturbation theory at an energy scale $k_B T_K \sim D e^{-1/(\nu J)}$, where $D$ is an energy cutoff and $\nu$ is the electron density of states. At energies below $T_K$, the impurity spin gets partially screened from $S_{\rm imp}=1$ to $S^*_{\rm imp} = \frac{1}{2}$ by the conduction electrons which, as a consequence, acquire a phase shift of $\delta_0 = \pi/2$. However, the residual moment, $S^*_{\rm imp}$, still couples ferromagnetically to the electron spin density. The ferromagnetic spin-$\frac{1}{2}$ Kondo model, complemented by an additional phase shift of $\delta_0 = \pi/2$  for the conduction electrons is thus expected to capture the properties of the Hamiltonian (\ref{Model}) at low energies. 

Although a ferromagnetic Kondo coupling is marginally irrelevant in the renormalization group (RG) sense, it results in important corrections to the Fermi liquid properties. Upon RG improvement, the effective ferromagnetic coupling acquires an energy dependence $\nu J_{\rm eff}(\varepsilon) = - 1/\log\left(k_B T^*_K/\varepsilon\right)$ for $\varepsilon \ll k_B T_K^*$ where one identifies $T_K^* \sim T_K$. This logarithmic energy  dependence 
is at the origin of the singular Fermi liquid properties of the underscreened
spin-$1$ Kondo model. Calculating the phase shift for the conduction electrons
in the Born approximation in $J_{\rm eff}$ one obtains, for example, in a
small magnetic field, $g\mu_B B \ll k_B T_K$, the spin-dependent correction to $\delta_0 = \pi/2$,
\begin{align} \label{PhaseShift}
\delta_\sigma  = \frac{\pi}{2} 
\left(1 + \sigma \frac{1}{2 \log\left(k_B T_K/g\mu_B B\right)} \right).
\end{align}

The focus of the present study are the waves within the Fermi sea that surge around the spin-$1$ impurity. 
A snapshot of these is provided by the equal-time correlator of the impurity spin and the conduction electron spin-density at position $x$,
\begin{align}
\chi(x,T) = \langle \vec{S}_{\rm imp}\,\vec{s}(x) \rangle.
\end{align}
We also consider the integrated correlator $\chi(T) = \int dx\, \chi(x,T)$. Using rotational invariance of the Hamiltonian (\ref{Model}), the latter can be expressed as an expectation value of the z-component of the impurity spin, $S_{\rm imp}^z$,  and total spin, $S_{\rm tot}^z = S_{\rm imp}^z + \int dx s^z(x)$, 
\begin{align} \label{SumRule}
\chi(T) = 3 \langle S_{\rm imp}^z S_{\rm tot}^z \rangle - 2.
\end{align}
This relation is particularly useful for a numerical evaluation of  $\chi(T)$ with the help of the NRG. 


{\em Method} --- Generally, the calculation of spatial correlations is a non-trivial task since
most of the methods used to investigate Kondo models are not able
to reproduce correlation functions. One of the most powerful numerical
methods, Wilson's NRG was believed to be unable to handle spatial correlations
for a long time. Only very recently it has been extended to have good
spatial resolution by one of the authors \cite{Borda07}. The approach
is similar to which was used to handle the two impurity 
problem with NRG \cite{SilvaWilkins96}. 
The key idea behind that method is to map the single channel problem onto
an effectively two channel case in order to have good spatial resolution
not only around the position of the impurity but around another freely chosen
point $x$ as well. For the technical details we refer the 
reader to Ref.~\onlinecite{Borda07}. In such a scheme the 
computation of the equal time correlation
function is rather simple: $\chi(x,T)$ appears to be a static
thermodynamic quantity which can be evaluated with a high precision.

{\em Results} --- The numerically evaluated correlator $\chi(x,T)$ for different Kondo couplings $J$ (and one-dimensional conduction electrons) is shown in Fig.~\ref{fig:equalt}a. It exhibits characteristic $2 k_F$ oscillations, where $k_F$ is the Fermi momentum. Very close to the impurity, $x k_F \ll 1$, as well as at distances corresponding to integer multiples of $\pi/k_F$ the spin of the conduction electrons is antiferromagnetically aligned with the impurity spin. In between, $\chi(x,T)$ changes its sign and shows ferromagnetic correlations. Upon excluding the immediate vicinity of the impurity, i.e., distances smaller than $\pi/k_F$, we find that the correlator collapses to an universal curve of the form  
\begin{align} \label{UniversalAsympt}
\lefteqn{\left.   \chi(x,T) \right|_{x > \pi/k_F} = \frac{1}{\xi_K} \times}
\\\nonumber& 
\left[
\mathcal{X}_{2 k_F}\left(\frac{x}{\xi_K},\frac{T}{T_K}\right) \cos\left(2 k_F x + 2 \delta_0\right) + \mathcal{X}_{un}\left(\frac{x}{\xi_K},\frac{T}{T_K}\right)
\right]
\end{align}
with two universal functions $\mathcal{X}_{2 k_F}$ and $\mathcal{X}_{un}$. Note that this is quite unlike the 
situation for the Friedel oscillations \cite{AffleckSaleur08}: in that case 
the non-universal part of the charge density oscillations
is much more extended in space. The additional factor $1/\xi_K$ in Eq.~(\ref{UniversalAsympt}) accounts for the correct dimensional units of $\chi(x,T)$. (Note that $\vec{s}(x)$ is a one-dimensional spin-density.) Moreover, we anticipated already the phase shift $\delta_0 = \pi/2$ that the electrons attain from screening of half of the impurity spin-$1$. In order to identify the two functions $\mathcal{X}_{2 k_F}$ and $\mathcal{X}_{un}$, we independently determined the Kondo temperature $T_K$. This was done by extracting the single particle phase shifts in the presence of a local magnetic field $B$ directly from the NRG spectra and fitting them to Eq.~(\ref{PhaseShift}).


The two universal functions $\mathcal{X}_{2 k_F}$ and $\mathcal{X}_{un}$ at zero temperature, $T=0$, are shown in Fig.~\ref{fig:equalt}b. Both decay algebraically as $1/x$ supplemented by logarithmic corrections for large distances, $x > \xi_K$. In the presence of a finite temperature, the decay changes from algebraic to exponential for $x>\xi_T$, where $\xi_T = \hbar v_F/(k_B T)$ is the thermal length, as it is shown in panel Fig.~\ref{fig:equalt}c for the amplitude of the oscillating part. 

It is expected that the long-distance behavior of $\chi(x,T)$ at $T \ll T_K$ is captured by an effective ferromagnetic spin-$\frac{1}{2}$ Kondo model. Assuming that at large distances $x \gg \xi_K$, the correlator is dominated by the unscreened part of the magnetic moment, $S^*_{\rm imp} = \frac{1}{2}$,
\begin{align}
\chi(x,T) \sim \langle \vec{S}^*_{\rm imp}\,\vec{s}(x) \rangle, \qquad {\rm for}\quad x \gg \xi_K
\end{align}
we can apply perturbation theory in the effective ferromagnetic coupling $J_{\rm eff}(\varepsilon)$ to calculate the asymptotic behavior. The amplitude of the $2k_F$ oscillating part starts in first order and the amplitude of the uniform part in second order in $J_{\rm eff}(\varepsilon)$. Its logarithmic energy dependence results in characteristic logarithmic corrections to power-law decay \cite{BarzykinAffleck98},
\begin{align} \label{UniversalLongAsympt}
\chi(x,T=0) =
\frac{\mathcal{C}}{x} \left[\frac{\cos\left(2 k_F x + 2 \delta_0\right)}{ \log (x/\xi_K)} - \frac{2}{  \log^2 (x/\xi_K)} \right]
\\\nonumber
{\rm for }\quad x \gg \xi_K.
\end{align}
Note that the antiferromagnetic correlations of $\chi$ observed at $x = n \pi / k_F$, with integer $n$, is an unambiguous signature of the additional phase shift $\delta_0 = \pi/2$ due to the partial screening of half of the impurity spin-$1$. The numerical determination of the coefficient $\mathcal{C}$ is handicapped by the fact that the NRG data  only yields reliable information on the spatial spin correlations up to $x/\xi_K \sim 10^3 $ such that the asymptotic regime is barely reached. In order to compare with Eq.~(\ref{UniversalLongAsympt}) nonetheless, we employ a two-parameter fit. In addition to the parameter $\mathcal{C}$ in (\ref{UniversalLongAsympt}), we allow the Kondo temperature extracted from a fit to the leading-logarithmic expression of the phase shift (\ref{PhaseShift}) to vary by an additional constant parameter, $\xi_K = \mathcal{A}\, \hbar v_F/(k_B T^{\rm phase\, shift}_{K})$. Adjusting the parameters $\mathcal{A}$ and $\mathcal{C}$ for a best fit to the long-distance tail, we obtain $\mathcal{C} = 0.07 \pm 0.008$.

\begin{figure}
\includegraphics[width=0.35\textwidth]{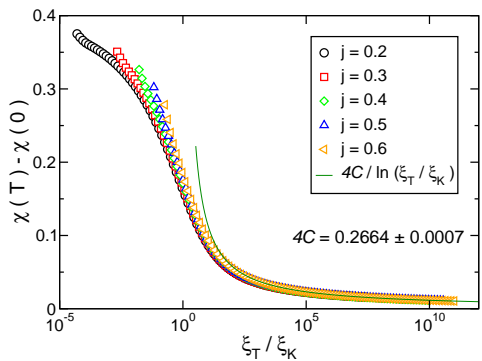}
\caption{(Color online) Temperature correction, $\chi(T)-\chi(0)$ to the integrated spatial spin correlator (\ref{SumRule}). At low temperatures $\xi_T \ll \xi_K$, it follows Eq.~(\ref{SumRuleTail}).}
\label{fig:SumRule}
\end{figure}

Within the numerical accuracy, we find the long-distance tail of $\chi(x,T)$
to be described by an universal coefficient $\mathcal{C}$ in
Eq.~(\ref{UniversalLongAsympt}) independent of the Kondo coupling $J$. On the
other hand, a renormalization group treatment of the effective ferromagnetic
Kondo model predicts the coefficient $\mathcal{C}$ to depend explicitly on the
effective bare ferromagnetic coupling \cite{BarzykinAffleck98}. The numerical
finding thus indicates that the low-energy model is characterized by a
universal dimensionless ferromagnetic Kondo coupling of the order of one. 

This is borne out by considering the integrated correlator $\chi(T)$ in Eq.~(\ref{SumRule}). At small but finite temperature, $T \ll T_K$, the correction $\chi(T) - \chi(0)$ is determined by the long-distance tail of the spin correlator $\chi(x,T=0)$ at {\em zero} temperature as, effectively, the temperature only cuts off the correlation tail at the distance of the thermal length $\xi_T$. From Eq.~(\ref{UniversalLongAsympt}) we thus obtain the asymptotics
\begin{align} \label{SumRuleTail}
\chi(T) - \chi(0) = \frac{4\, \mathcal{C}}{\log( \xi_T/\xi_K)}
\quad {\rm for}\quad \xi_T \gg \xi_K.
\end{align}
This allows for an alternative determination of the coefficient $\mathcal{C}$
by computing $\chi(T)$ with NRG. The result is shown in
Fig.~\ref{fig:SumRule}. It allows a fit to Eq.~(\ref{SumRuleTail}) over more
than seven decades in $\xi_T/\xi_K$ yielding a more accurate value for
$\mathcal{C} = 0.0666 \pm 0.0002$.  Whereas $\mathcal{C}$ is found to be universal, the value $\chi(T=0)$ depends on the Kondo coupling $J$ (not shown).

To summarize, we have demonstrated that the profile of the spatial spin correlations, $\chi(x,T)$, around an underscreened spin-$1$ is markedly different from the one of a fully screened spin-$\frac{1}{2}$. The correlations are more pronounced and decay at $T=0$ as $1/x$ in $d=1$ with logarithmic corrections, and they change sign 
as a function of distance with ferromagnetic correlations at $x k_F/\pi = n+\frac{1}{2}$ with integer $n$. The long-distance behavior is universal and can be explained in terms of an effective ferromagnetic Kondo model.

{\em Acknowledgments ---}
This work was supported by the DFG through SFB 608. 
L.B. acknowledges the financial support of the Alexander von Humboldt
Foundation, J\'anos Bolyai Foundation and Hungarian Grants OTKA
through projects T048782 and K73361.


\begin{thebibliography}{99}


\bibitem{Hewson} A.C. Hewson, {\it The Kondo Problem to Heavy Fermions}
		   (Cambridge University Press, 1993).
		   
\bibitem{Ishii78} H. Ishii, J. Low Temp. Phys. {\bf 32}, 457 (1978).		   

\bibitem{BarzykinAffleck98} V. Barzykin and I. Affleck, Phys. Rev. B {\bf 57}, 432 (1998).

\bibitem{Affleck01}
I. Affleck and P. Simon, Phys. Rev. Lett. {\bf 86} 2854 (2001) 


\bibitem{Sorenson05}
E.S. S{\o}rensen and I. Affleck, Phys. Rev. Lett. {\bf 94}, 086601 (2005).

\bibitem{Borda07} L. Borda, Phys. Rev. B {\bf 75}, 041307(R) (2007).

\bibitem{AffleckSaleur08} I. Affleck, L. Borda, H. Saleur,
Phys. Rev. B {\bf 77}, 180404(R) (2008).



\bibitem{MehtaZarand05}P. Mehta, N. Andrei, P. Coleman, L. Borda, and G. Zar\'and,
Phys. Rev. B {\bf 72}, 014430 (2005).

\bibitem{KollerMeyer05}W. Koller, A.C. Hewson, and D. Meyer, Phys. Rev. B {\bf
    72}, 045117 (2005).
    

\bibitem{orsi} O. \'Ujs\'aghy, A. Zawadowski, and B.L. Gyorffy,
  Phys. Rev. Lett. {\bf 76}, 2378 (1996).
  
\bibitem{rok} R. \v{Z}itko, R. Peters, and Th. Pruschke, e-print arXiv:0809.0759.

    
\bibitem{PustilnikGlazman04}

M. Pustilnik and L.I. Glazman, Phys. Rev. Lett. {\bf 87}, 216601 (2001).

\bibitem{PosazhennikovaColeman05}A. Posazhennikova and P. Coleman,
  Phys. Rev. Lett. {\bf 94}, 036802 (2005).
  
  


\bibitem{SilvaWilkins96} J. B. Silva, W. L. C. Lima, W. C. Oliveira, J. L. N. Mello,
  L. N. Oliveira, and J. W. Wilkins, Phys. Rev. Lett. {\bf 76}, 275 (1996)

\end{thebibliography}
\end{document}